
\def\ifundefined#1{\expandafter\ifx\csname
	#1\endcsname\relax}

\newcount\eqnumber \eqnumber=0
\def\beq{ \global\advance\eqnumber by 1 $$ }
\def\eeq{ \eqno(\the\eqnumber)$$ }
\def\label#1{\ifundefined{#1}
\expandafter\xdef\csname #1\endcsname{\the\eqnumber}
\else\message{label #1 already in use}\fi}
\def\(#1){(\csname #1\endcsname)}

\newcount\refno \refno=0
\def\[#1]{\ifundefined{#1}\advance\refno by 1
\expandafter\xdef\csname #1\endcsname{\the\refno}
\fi[\csname #1\endcsname]}
\def\refis[#1]{\item{\csname #1\endcsname.}}

\def\section#1{\vskip12pt\centerline{\twelvecp#1}\vskip8pt}

\footline={\ifnum\pageno=1\else\hfil\folio\hfil\fi}
\def\makefootline{\baselineskip=1.6cm\line{\the\footline}}

\newfam\scrfam
\batchmode\font\tenscr=rsfs10 \errorstopmode
\ifx\tenscr\nullfont
	\message{rsfs script font not available. Replacing with calligraphic.}
\else	\font\sevenscr=rsfs7
	\font\fivescr=rsfs5
	\skewchar\tenscr='177 \skewchar\sevenscr='177 \skewchar\fivescr='177
	\textfont\scrfam=\tenscr \scriptfont\scrfam=\sevenscr
	\scriptscriptfont\scrfam=\fivescr
	\def\scr{\fam\scrfam}
	\def\cal{\scr}
\fi
\newfam\frakfam
\batchmode\font\twelvefrak=eufm10 scaled\magstep1 \errorstopmode
\ifx\twelvefrak\nullfont\def\frak{\it}
	\message{Euler fraktur not available. Replacing with italic.}
\else	\font\tenfrak=eufm10 \font\sevenfrak=eufm7 \font\fivefrak=eufm5
	\textfont\frakfam=\tenfrak
	\scriptfont\frakfam=\sevenfrak \scriptscriptfont\frakfam=\fivefrak
	\def\frak{\fam\frakfam}
\fi
\newfam\msbfam
\batchmode\font\twelvemsb=msbm10 scaled\magstep1 \errorstopmode
\ifx\twelvemsb\nullfont\def\Bbb{\bf}
	\message{Blackboard bold not available. Replacing with boldface.}
\else	\catcode`\@=11
	\font\tenmsb=msbm10 \font\sevenmsb=msbm7 \font\fivemsb=msbm5
	\textfont\msbfam=\tenmsb
	\scriptfont\msbfam=\sevenmsb \scriptscriptfont\msbfam=\fivemsb
	\def\Bbb{\relax\ifmmode\expandafter\Bbb@\else
 		\expandafter\nonmatherr@\expandafter\Bbb\fi}
	\def\Bbb@#1{{\Bbb@@{#1}}}
	\def\Bbb@@#1{\fam\msbfam\relax#1}
	\catcode`\@=\active
\fi
\font\fourteenrm=cmr10 scaled\magstep 2
\font\twelvecp=cmcsc10 scaled\magstep 1
\baselineskip=15pt
\def\R{{\Bbb R}}
\def\C{{\Bbb C}}

\def\sE{{\rm sign}(E)}
\def\ord{{\rm ord}\,}
\def\eps{\varepsilon}
\def\st{c_{S}(A,B)}
\def\psdo{$\Psi${\rm DO}}
\def\trrig{{\rm Tr}}
\def\Trc{{\rm Tr}_C}
\def\tr{\;{\rm tr}\,}
\def\res{{\rm Res}\,}
\def\sym{{\rm sym}\,}
\def\glres{{\frak gl}_{\rm res}}
\def\glpsdo{\Psi{\frak gl}_{\rm res}}
\def\intall{\int_{\R^3} {{d^3p}\over{(2\pi)^3}}\int_D d^3x \,{\rm tr}}
\def\intalln{\int_{\R^n} {{d^np}\over{(2\pi)^n}}\int_D d^nx\,{\rm tr}}
\def\intdelta{\int_{|p|\leq\delta}
        {{d^3p}\over{(2\pi)^3}}\int_D d^3x \,{\rm tr}}
\def\*{\partial}
\def\limd{\lim_{\delta\to 0^+}}

\noindent G\"oteborg ITP 94-23

\noindent{\tt hep-th/9410016 }
\vskip1in

\centerline{{\fourteenrm Schwinger Terms}}
\centerline{{\fourteenrm and}}
\centerline{{\fourteenrm Cohomology of Pseudodifferential
		Operators}}

\vskip.6in

\centerline{Martin~Cederwall, Gabriele~Ferretti, Bengt~E.W.~Nilsson
	and Anders~Westerberg}
\vskip.5cm
\centerline{\sl Institute of Theoretical Physics}
\centerline{\sl Chalmers University of Technology}
\centerline{\sl and G\"oteborg University}
\centerline{\sl S-412 96 G\"oteborg, Sweden}

\vskip 1in
\centerline{\bf Abstract}
\noindent We study the cohomology of the Schwinger term arising in
second quantization of the class of observables belonging to the
restricted general linear algebra. We prove that,
for all pseudodifferential operators in 3+1 dimensions of this type,
the Schwinger term
is equivalent to the ``twisted'' Radul cocycle, a modified version
of the Radul cocycle arising in non-commutative differential
geometry. In the process we also show how the ordinary Radul cocycle for any
pair of pseudodifferential operators in any dimension
can be written as the phase space integral of the star commutator
of their symbols projected to the appropriate asymptotic component.
\vfill
\catcode`\@=11
\vbox{\hbox to\hsize{E-mail: \tt tfemc@fy.chalmers.se\hfill}
	\vskip-4pt
\hbox to\hsize{\phantom{E-mail: }\tt ferretti@fy.chalmers.se\hfill}
	\vskip-4pt
\hbox to\hsize{\it\phantom{E-mail: }\tt tfebn@fy.chalmers.se\hfill}
	\vskip-4pt
\hbox to\hsize{\it\phantom{E-mail: }\tt tfeawg@fy.chalmers.se\hfill}}
\catcode`\@=12

\eject

\vsize=21.5cm
\hsize=15cm
\voffset=.5cm
\hoffset=1cm

\section{1. Introduction}

Current algebras play an important role in many quantum field theories.
Historically,
they were introduced in an attempt to describe hadronic processes.
The hope was that the relevant physics would be captured by a restricted
set of operators, the currents, satisfying linear commutation relations
among themselves, and by a hamiltonian, bilinear in the currents, describing
their time evolution. Even after the advent of QCD as the ``microscopic''
theory of strong interactions, physicists have often used current algebra
techniques in the kinematical regions where the fundamental theory becomes
intractable.

When seen from the point of view of a more fundamental theory, the currents are
interpreted as composite operators in terms of the elementary fields,
e.g.~bilinears in some fermionic matter field.
Often, at the quantum level, the naive conservation laws and commutation
relations of the currents have to be modified by the addition of extra terms.
In particular, when they spoil the conservation laws of some
classically conserved current,
these terms are referred to as anomalies. These anomalies
are of crucial importance for the physical applications of the algebra; for
global algebras they are known to determine e.g.~the decay rate
of $\pi$ mesons, while for local algebras one is confronted by  unitarity
problems if the  extension cannot be eliminated
by choosing the particle content of the theory properly.

When appearing directly in the current--current commutation relations, these
terms are also referred to as Schwinger terms because originally such terms
were introduced by Schwinger in the context of QED [Sc-59]. From the point of
view of the fundamental theory, they should be generated by
the regularization procedure needed to make the current a well-defined
composite operator. Their effect on the commutation
relations can be understood in terms of Lie algebra cohomology as giving a
certain central or abelian (perhaps
even non-abelian) extension of the naive current algebra.

We will only consider the case where the currents
are bilinear in some fermionic field. In 1+1 dimensions we know from many
thoroughly studied examples
(e.g.~affine Kac--Moody algebras [Ba-71, Ka-67, Mo-67]) that normal
ordering suffices to make such currents well defined
and that, in general, central extensions are generated. In 3+1 and higher
dimensions the situation changes dramatically in that normal ordering
alone is not enough to render the bilinear expressions for the currents
well defined. However, although in perturbation theory
a (wave function) renormalization that successfully eliminates
this problem can be implemented,
it is still not understood how to define the currents in a completely regular
fashion.

As we will discuss extensively below, some of these concepts can be
rigorously formulated using the language of second quantization. In
particular, to any observable in the one-particle Hilbert space, one can
associate a fermionic bilinear acting in some Fock space. From this point
of view, the ordinary currents are thought of as second-quantized
multiplicative operators, and in dimensions higher than 1+1 they require
further regularization in addition to normal ordering.

It is of interest to isolate the observables for which normal ordering
is sufficient even in higher dimensions. These form what is known as the
restricted general linear algebra $\glres$ of the one-particle Hilbert space.
In particular, we will show that it is possible to characterize these
operators explicitly by considering only pseudodifferential operators
(\psdo s).
This can hardly be regarded as a loss of generality, since all the operators
of interest in physics can be regarded as \psdo s of some kind.
The real restriction is in considering only operators in $\glres$.
Nevertheless, the study of $\glres$ is of great interest for at least three
independent reasons:
\item{i)} The approach works in (1+1)-dimensional spacetime, in the sense
that normal ordering in this case suffices to regularize most operators. In
particular, all affine Kac--Moody algebras can be understood in this way.
\item{ii)} In higher dimensions $\glres$ represents a simple subclass of
operators that can be studied very explicitly, still displaying non-trivial
properties such as the
presence of Schwinger terms in their commutators. Any future understanding of
representation theory of higher-dimensional current algebras must eventually
agree with the results obtained for this subclass.
\item{iii)} $\glres$ may actually be of crucial importance in developing
the representation theory mentioned above. It has recently been proposed by
Mickelsson [Mi-93] that the elements of $\glres$ should be used as regularized
versions of the more singular operators one is actually interested in.
$\glres$ should play a similar role in the study of the generalization of
higher-dimensional current algebras recently discovered in [Ce-94, Fe-94].

Normal ordering of the second quantized \psdo s in $\glres$ generates
Schwinger terms which appear as two-cocycles of the underlying Lie algebra.
As such, they define a central extension $\widehat\glres$ of $\glres$.
However, when discussing \psdo s one finds that the requirement
of making them smooth at zero momentum introduces a regulating function,
i.e.\ the Schwinger term becomes regularization dependent. This is an
unwanted feature of the procedure, and it is crucial
to find a way to extract the cohomological information,
or, in other words, to relate the cohomology class of the Schwinger term to
one of the known cohomologies in the space of \psdo s. How this can be done
is one of the two main results of our paper:

\item{$\bullet$ }{\it The Schwinger term for {\rm \psdo}'s in
{\rm $\glres$} lies in
the same cohomology class as the so-called ``twisted'' Radul cocycle
{\rm [Mi-94]}, a slightly modified version of the well-known Radul cocycle
used in non-commutative differential geometry.}

\noindent Our second main result (that will actually be proven
first)
is not related in any way to the
structure of $\glres$, but is a general result on the cohomology
of \psdo s:

\item{$\bullet$ }{\it In any number of dimensions $n$, the Radul cocycle of
two arbitrary {\rm \psdo}'s (not necessarily in {\rm $\glres$}) can be written
as the
integral over all phase space of their commutator projected onto the
component with asymptotic behavior $|p|^{-n}$.}

The paper is organized as follows. After some introductory material
on second quantization and Schwinger terms in sections~2 and 3, respectively,
we introduce \psdo s in sect.~4. These short
sections cover only well-known material and are added primarily in an attempt
to make the paper easier to read and to a certain extent self-contained.
In sect.~5 we prove that the Radul
cocycle can be expressed as a commutator as stated
above. In sect.~6 we characterize the \psdo s that
belong to $\glres$ and use this characterization in sect.~7
to relate the Radul cocycle to the Schwinger term.
Some additional remarks are added in sect.~8
and we mention a few cases where our results are directly relevant,
namely, affine Kac--Moody algebras [Ba-71, Ka-67, Mo-67] in 1+1 dimensions,
Mickelsson--Faddeev--Shatashvili algebras in 3+1 dimensions [Fa-84a, Fa-84b,
Mi-83]
and a recently proposed extension of the algebra of maps from an
$n$-dimensional manifold into a semisimple Lie algebra [Ce-94, Fe-94].
We plan to return to these examples, particularly the last one,
in a future publication. For some recent results in this area,
see~[Ba-93, Ca-94, La-94a].

\section{2. From first to second quantization}

Consider a particle moving in Minkowski space $\R^{(n,1)}$.
In quantum mechanics, the dynamics of such a particle is specified by
giving the time evolution of its wave function $\Psi : \R^n \rightarrow V$
up to an overall complex phase. Here $V$ denotes the $M$-dimensional complex
vector space describing the other degrees of freedom of the particle,
namely spin and color. Throughout this paper we will only consider the case
of half-integral spin. ``Color'' here simply means any internal symmetry the
system
may have.

To be specific, we will in sect.~6 restrict our attention to
(3+1)-dimensional Weyl spinors transforming
in the
fundamental representation of the color group ${\frak su}(N)$. The wave
function $\Psi$
is then valued in the ($M=2N$)--dimensional complex vector space
$V=\C^2_{\rm spin}\otimes\C^N_{\rm color}$. The restriction to
Weyl spinors in 3+1 dimensions will be made because,
on the one hand, this is the most interesting case due to its direct
connection to chiral gauge theory and, on the other hand, it is simple enough
to allow explicit calculations, yet general enough to display all the issues
we want to discuss. However, the analysis can be repeated for
particles with other spins~[La-91]. In any case, all that is said in this
and in the
following section depends only on the fermionic nature of the matter field and
not on the specific representation or spacetime dimension.

We must of course restrict ourselves to square-integrable
wave functions forming the first-quantized Hilbert space ${\cal H}$.
At the level of quantum mechanics, the observables are described by
self-adjoint operators $A=A^\dagger$ acting
on ${\cal H}$. We do not need to worry about
questions of domain in the first-quantized Hilbert space since all operators
of interest to us are bounded.

However, as is well known from the early days of quantum mechanics this picture
is inadequate if we want to describe the relativistic dynamics of elementary
particles because the energy $E$ of the free particle is not bounded below and
creation/annihilation processes cannot be described. In mathematical
terms, ${\cal H}$ carries a
representation of the algebra of observables (to which $E$ belongs) that
is not highest (actually lowest) weight.
The solution to this problem in the Hamiltonian formulation is also well known;
precisely because of its privileged status in defining the lowest weight, one
uses the first-quantized energy operator $E$ to define a polarization, i.e.~a
splitting of the Hilbert space into non-negative and negative energy spaces
${\cal H}={\cal H}_+\oplus {\cal H}_-$.
Then one introduces a new Hilbert space ${\cal F}$ (the
Fock space), a lowest weight vector $|0\rangle \in {\cal F}$ (the vacuum),
and a set of operators acting on ${\cal F}$, $a(\Psi)$ and $a^\dagger(\Psi)$
(the
annihilation and creation operators, respectively),
satisfying $a(\Psi)|0\rangle=0$ if $\Psi\in {\cal H}_+$ and $a^\dagger
(\Psi)|0\rangle=0$ if
$\Psi\in {\cal H}_-$. Since
we
are only considering fermionic fields, the spin--statistics theorem requires
that these operators satisfy the anti-commutation relations
$\{ a(\Psi_1), a^\dagger(\Psi_2)\}=\langle\Psi_1|\Psi_2\rangle$.

With these assumptions, the Fock space carries an irreducible representation of
the canonical anticommutation relations. One can then represent the algebra of
observables in the Fock space, i.e.\ second quantize the theory, as follows.
Consider a basis of eigenfunctions $\{ \psi_n \} \in {\cal H}$ of $E$. Here
$n$ is a generic multi-index labeling the elements of the basis and we
write symbolically $n\geq 0$ iff $\psi_n \in {\cal H}_+$ and $n<0$ iff
$\psi_n\in {\cal H}_-$ . Also, for
the sake of brevity, we define $a_n = a(\psi_n)$,
$a^\dagger_n = a^\dagger(\psi_n)$ and $A_{mn}=\langle\psi_m|A|\psi_n\rangle$.
The representation of $A$ in the second-quantized Fock space is then given by
the operator $\hat A = \sum_{mn}A_{mn} :a^\dagger_m a_n:$, where the colons
represent the normal ordering necessary to ensure that the operators have zero
vacuum expectation value. One way
to realize the normal ordering is by setting
\beq
     :a_m^\dagger a_n:\,= \cases{-a_n a_m^\dagger& for $n$ and $m <0,$\cr
                               a_m^\dagger a_n & otherwise.\cr}
\eeq
If such a representation exists, it is manifestly unitary
and lowest weight, i.e. $\hat E$ is bounded below by the vacuum energy
$\hat E|0\rangle=0$.

What can go wrong in going from first to second quantization? In other words,
how do we make sure that $\hat A$ exists? The condition to check is that
$\hat A $ creates states of finite norm out of the vacuum, i.e. that
$\| \hat A|0\rangle\| <\infty$. This norm can be computed explicitly as
\beq
     \|\hat A|0\rangle\|^2 =
     \langle 0|\hat A^\dagger \hat A|0\rangle =
		\sum_{m\geq 0,n<0}A^*_{mn} A_{mn}
     = {1\over 8}\trrig{([\sE,A]^\dagger [\sE,A])},
	\label{normA}
\eeq
where $\sE = \pm 1$ on ${\cal H_\pm}$. Hence, $\hat A$ is well defined
iff the square of $[\sE,A]$ has finite trace in ${\cal H}$. Operators whose
square have finite trace are known as Hilbert--Schmidt (HS) operators. With
respect to
the polarization ${\cal H}={\cal H}_+ \oplus {\cal H}_-$ an arbitrary operator
$A$ and, in particular, the sign of the energy operator can be written as
\beq
     A=\pmatrix{A_{++} & A_{+-} \cr A_{-+} & A_{--} \cr},\quad
     \quad \sE=\pmatrix{1&0\cr 0&-1\cr}.
\eeq
Requiring the commutator $[\sE,A]$ to be HS is equivalent to requiring
that the off-diagonal blocks of $A$ be separately HS.
In order for the algebra of observables to close under this property, one
must also require that the elements be bounded operators.
This algebra is called
the restricted general linear algebra $\glres$:
\beq
    \glres=\big\{A : {\cal H}\to {\cal H}\,\,\hbox{ bounded}
     \,\,\,|\,\,\,[\sE, A]\,\,
     \hbox{Hilbert--Schmidt}\big\}. \label{defofglres}
\eeq

At this point, we would like to make a short digression on the precise
definition of the trace in order to avoid confusion. Similar comments can be
found in [La-94b]. A arbitrary bounded linear operator $S$ on a Hilbert space
${\cal H}$ is said to be trace class (see e.g.~[Pr-86, Si-79]) if its action
on an arbitrary vector
$\Phi\in
{\cal H}$ can be written as
\beq
     S\Phi= \sum_k \lambda_k \langle\psi_k|\Phi\rangle\chi_k,
\eeq
where $\{ \psi_n \}$ and $\{\chi_m \}$ are two orthonormal Hilbert bases of
${\cal H}$ and $\sum |\lambda_n| <\infty$. (Notice that it may not be possible
to choose $\psi_n\!=\!\chi_n$ if $S$ does not have a complete set of
eigenvalues.)
For trace class operators the trace is defined to be
\beq
      \trrig(S)=  \sum_k \lambda_k\langle\psi_k|\chi_k\rangle.
      \label{rigoroustrace}
\eeq
This series is obviously absolutely convergent since
$|\langle\psi_k|\chi_k\rangle|\leq 1$. Such a
trace is basis independent; the two families $\{ \psi_n \}$ and $\{
\chi_m \}$ define the operator, not the trace.

In our applications, however, the Hilbert space comes with a
polarization and the kind of trace that we need is
\beq
    \Trc S= \trrig  \pmatrix{S_{++}&0\cr
                                  0     & S_{- -}\cr }
    = {1\over 2} \trrig (S + \sE\, S\; \sE), \label{condtrace}
\eeq
with the traces in the middle and on right hand side defined as in
\(rigoroustrace). By considering $S=[\sE,A]^\dagger [\sE,A]$, where
$A\in\glres$, we see that we could use $\Trc$ instead of $\trrig$ in
eq.~\(normA).
Clearly, if $S$ is trace class the two definitions coincide. However,
the trace $\Trc$ is convergent for a larger class of operators (called
``conditionally trace class'' in [La-94b]) since
the combination $S + \sE S\; \sE$ projects out the potentially too singular
off-diagonal terms. The price one has to pay is that the definition of
$\Trc$ depends on the choice of polarization. Obviously, the projection
$S\mapsto (1/2)(S + \sE S \sE)$
is idempotent, and therefore, whereas $\trrig(S) =
(1/2)\trrig(S + \sE S \sE)$ only for truly trace class operators,
$\Trc (S) = (1/2)\Trc (S + \sE S \sE)$ holds for the whole class of
conditionally trace class operators.

Unfortunately, the operators of ordinary quantum mechanics, in general, do not
admit a second-quantized representation like the one described above,
i.e.~they do not belong to $\glres$, and one therefore needs to
renormalize the vacuum expectation values~[Mi-88, Fu-90, Pi-87, Pi-89].
As an illustrative example, consider
a smooth function $f(x)$ with compact support and define the multiplicative
operator $(F\Psi)(x) \equiv f(x)\Psi(x)$. It is readily checked that
$\hat F|0\rangle$ has finite norm, i.e.~belongs to $\glres$, only in 1+1
dimensions. Nevertheless, as mentioned in the introduction, there are many
reasons for
looking at $\glres$, perhaps the most important one being
that this allows one to obtain rigorous results for a specific class of
observables that will eventually have to be matched by any
more general method.

\section{3. The Schwinger term as a two-cocycle}

As mentioned in the introduction,
one of the subtleties arising in quantum field theory
is the appearance of $c$-number terms in the commutation relations of various
operators, so-called Schwinger terms. A simple
example of such a term is the one present in the commutator
between the space and time components of the normal-ordered electromagnetic
current $J_\mu (\vec x,t)$ (for, say, QED). The naive expectation that
$[J_0 (\vec x,t),J_k (\vec y,t)]=0$ is frustrated by the fact that current
conservation would then require $J_0$ to vanish. Schwinger postulated the
appearance of the derivative of a $\delta$-function on the right hand side
of the equation, that, vanishing upon integration,
does not spoil the definition
of electric charge: $[J_0 (\vec x,t), J_k (\vec y,t)]= {\rm const}\times
i\partial_k \delta(\vec x - \vec y)$. That this term actually arises can be
proven rigorously in 1+1 dimensions by taking the current to be a
normal-ordered fermionic bilinear and using point-splitting regularization.

The advantage of restricting ourselves to the operators in $\glres$ is
that the same rigorous calculations can be straightforwardly generalized to
arbitrary dimensions, if only for a very restricted class of operators. In
fact, at this abstract level, nothing depends on the dimension of spacetime,
i.e. on the particular choice of ${\cal H}$. Let us thus consider $A, B\in
\glres$ and set
\beq
     [\hat A, \hat B] = \widehat{[A,B]} -{1\over2} \st.
\eeq
(The factor $-1/2$ is inserted for later convenience.)
By taking the vacuum expectation value of both sides, and using the fact that
$\langle 0|0\rangle = 1$ and that $\langle 0|\widehat{[A,B]}|0\rangle = 0$
we obtain the Schwinger term
\beq\eqalign{
    \st =& -2\langle0|[\hat A, \hat B]|0\rangle\cr =&-{1\over 4}
         \trrig\bigg(\sE\big[[\sE,A],[\sE,B]\big]\bigg) \cr =&
        -{1\over 2}
         \trrig\bigg(\sE[\sE,A][\sE,B]\bigg) \cr = &
         \Trc\big([\sE,A]B\big).\cr}
	\label{schwingerterm}
\eeq
The traces are convergent precisely because of the HS property that we have
assumed for the operators $A$ and $B$. Moreover, the Schwinger term in
\(schwingerterm) turns out to be a two-cocycle of the algebra $\glres$
defining a non-trivial central extension known as
$\widehat{\glres}$.

Let us at this point recall some basic elements of Lie algebra cohomology in
order to keep our discussion self-contained.
For an extensive discussion of Lie algebra cohomology and its relation to
quantum field theory we refer the reader to e.g.~[Ka-90, Ki-76, Mi-89].
Given an abstract Lie algebra ${\cal L}$, an $n$-cochain with values in $\C$
is defined as an anti-symmetric $n$-linear map
$c^n:{\cal L}\wedge{\cal L}\wedge..\wedge{\cal L}\rightarrow\C$. We denote
the vector space of such $n$-cochains by $C^n=C^n({\cal L},\C)$.
The coboundary operator $\delta:C^n\rightarrow C^{n+1}$ is defined by
\beq
\delta c^n(x_1,x_2,...,x_{n+1})
 =\sum_{i<j}(-1)^{i+j+1}
c^n([x_i,x_j],x_1,...,\hat{x}_i,...,\hat{x}_j,...,x_{n+1}),
\label{coboundarydef}
\eeq
where a caret indicates an absent argument. In particular,
\beq\eqalign{
\delta c^1(x_1,x_2)&=c^1([x_1,x_2]), \cr
\delta c^2(x_1,x_2,x_3)&=c^2([x_1,x_2],x_3)+c^2([x_1,x_3],x_2)
	-c^2([x_2,x_3],x_1).\cr}
\eeq
The basic property of $\delta$ is its nilpotency, i.e $\delta^2=0$.
Cochains such that $\delta c =0$ are called cocycles (or closed cochains),
and cocycles of the form $c=\delta \lambda$ are called coboundaries (or exact
cochains). The abelian groups obtained by
considering linear combinations of cocycles modulo coboundaries define the
Lie algebra cohomology of ${\cal L}$.
The only application of Lie algebra cohomology that we will need in this
paper is
that the second cohomology group $H^2({\cal L},\C)$ describes the possible
central extensions of ${\cal L}$. Namely, on the vector space
${\cal L}\oplus\C$ the commutator
\beq
[(x,\xi),(y,\eta)] = ([x,y],c(x,y))
\eeq
defines a Lie algebra $\hat{\cal L}$
(i.e.~satisfies the Jacobi identities)
if and only if $c(x,y)$ is a two-cocycle.
Furthermore, two two-cocycles define isomorphic Lie algebras
if their difference is a coboundary. An algebra $\hat{\cal L}$ obtained in this
way, a central extension of ${\cal L}$ by $\C$, is thus specified by
an element of the second Lie algebra cohomology group $H^2({\cal L},\C)$.

Comparing with the definitions above, it is easily checked that the
Schwinger term \(schwingerterm) is in fact a non-trivial two-cocycle
sometimes also referred to as the Lundberg cocycle [Lu-76].
Understanding the explicit form of such terms and
their relation with other kinds of
cohomologies, namely those that arise in the study of pseudodifferential
operators (\psdo s), will be the scope of most of the remainder of this paper.

\section{4. Basic facts about pseudodifferential operators}

In order to keep the paper self-contained we present here some well-known
facts about pseudodifferential operators (\psdo s) that will be needed
later on. We only give the basic results without proofs and refer the reader
to e.g.~[H\"o-85, La-89, Ta-81, Va-93] for more detailed discussions.

Consider the Hilbert space ${\cal H}={\rm L}^2(\R^n)\otimes\C^M$ of
square integrable functions $\psi: \R^n \to \C^M$ of $x$. The
\psdo\  $S$ acting on ${\cal H}$ is defined by
\beq
S\psi(x)=\int e^{ix\cdot p}s(x,p)\tilde\psi(p){{d^np}\over{(2\pi)^n}},
	\label{psdodef}
\eeq
where $\tilde\psi(p)=\int e^{-ix\cdot p}\psi(x)d^nx$ is the Fourier
transform of $\psi$ and $s : \R^n\times \R^n \to {\frak gl}(M,\C)$ is a smooth
function assumed to have compact support in $x$ and at most polynomial growth
in $p$. The function $s(x,p)$ is called the symbol of $S$ which we write as
$\sym(S)=s$.

A \psdo\ $S$ (or its symbol $s$) is said to be of order $m$, written as
$\ord(S)=m$, if it has a leading asymptotic behavior for large $|p|$ of the
kind $s(x,p) = {\cal O}(|p|^m)$ uniformly in $x$.
Here we will only be concerned with \psdo s
of integral order. A \psdo\ whose symbol decreases faster
than any power of $p$ is called infinitely smoothing.
Two \psdo s $S$ and $R$ are said to be equivalent if they differ
by an infinitely smoothing operator. We will denote such an equivalence by
$S\approx R$ for the operators, or by $s\approx r$ for their symbols.

The importance of this equivalence relation is that it allows for the
introduction of asymptotic expansions; consider the sequence
$\{s_k (x,p), k\leq m\}$, where $s_k$ is a smooth
symbol of order $k$. A symbol $s$ of order $m$ is said to have the
asymptotic expansion
\beq
  s(x,p) \approx \sum_{k\leq m}s_k (x,p)
  \label{asymptoticexpansion}
\eeq
if, for each integer
$r\leq m$,
\beq
\ord\bigg( s(x,p)-\sum_{k=r}^m s_k(x,p)\bigg) =r-1.
\eeq
It is often most convenient to assume that the symbols $s_k$ in
the asymptotic expansion \(asymptoticexpansion) are homogeneous of degree
$k$ in $p$ for $|p|>\delta$ and smooth everywhere:
\beq
       s_k(x,\lambda p) = \lambda^k s_k(x,p)\quad\hbox{for}\quad
       \lambda>1,\hbox{~~and~~} |p|\geq \delta>0.
\eeq
This does not represent
a loss of generality, since any \psdo\ has such an asymptotic expansion.
The necessity of imposing $|p|\geq\delta$ arises from the fact that a
homogeneous function is not, in general, smooth at the origin;
in this sense, $\delta$ should be thought of as an infrared regulator
to be taken to zero at the end.

Any asymptotic expansion \(asymptoticexpansion)
defines the symbol of a \psdo\ up to an
infinitely smoothing operator and we can therefore use the same equivalence
sign ``$\approx$'' between two asymptotic expansions.
One way to convince oneself that this is true is to introduce a $C^\infty$
function $\phi : \R_+ \rightarrow \R$ such that $\phi(t)=0$ for $t<1/2$
and $\phi(t)=1$ for $t>1$, and set
\beq
    s(x,p) =\sum_{r\geq 0} \phi(|p|/(1+r)) s_{m-r}(x,p).
    \label{expansionofs}
\eeq
It can be proven that $s(x,p)$ is the symbol of a \psdo\ of order $m$.
Although there is a lot of arbitrariness in the choice of $s(x,p)$
it should be evident
that two such symbols can only differ by an infinitely smoothing operator.
Note that the regulating
function $\phi$ in the series \(expansionofs) for $s(x,p)$ above has the effect
of truncating the series for any given value of $|p|$ to a finite
number of terms,
and that the number of terms grows with increasing $|p|$.

The basic operation in symbol calculus is the star product,
corresponding to the (noncommutative) multiplication of
operators on Hilbert space. In other words, the star product of the symbols
of two operators $S$ and $R$ is defined as the symbol of the composite
operator:
\beq
      \sym(S)*\sym(R) \approx \sym(SR).
\eeq
The asymptotic expansion of the star product of two symbols is
\beq
     (s*r)(x,p)\approx \sum_{k=0}^\infty
       {{(-i)^k}\over{k!}}
       {{\partial^k s}\over{\partial p_{\mu_1}\cdots
       \partial p_{\mu_k}}}\,
       {{\partial^k r}\over{\partial x^{\mu_1}\cdots
       \partial x^{\mu_k}}},\label{stardef}
\eeq
and we may formally write
\beq
        *= \exp(-i {{\buildrel\leftarrow\over\partial}\over
       {\partial p_\mu}}{{\buildrel\rightarrow\over\partial}\over
       {\partial x^\mu}}).
\eeq
Note from the first term in the expansion \(stardef) that $\ord(SR)=
\ord(S)+\ord(R)$.
Although we have not explicitly inserted one in \(stardef),
a regulator is needed if one, as we do here, wants
to deal with smooth symbols only. Consequently, \(stardef) defines
such a smooth function only up to an infinitely smoothing operator.

The asymptotic behavior of the symbol also determines whether
the corresponding operator is bounded, HS or trace class; in any dimension
$n$, $S$ is bounded iff $\ord(S) \leq 0$, HS iff
$\ord(S) < -(n/2)$ and trace class iff $\ord(S) < -n$, the last two
inequalities being in the strict sense.  For a trace class \psdo\ one could,
of course, compute the trace as in def.~\(rigoroustrace), which by Fourier
analysis would lead to
\beq
        \trrig(S) = \intalln\,s. \label{naivetrace}
\eeq
There are, however, a couple of problems with this expression. One is
that it is not well defined on the equivalence classes of \psdo s;
for instance, $\trrig e^{-|\Delta|}\neq 0$. This means, for example,
that one should be careful in using asymptotic expressions like
\(stardef) inside this trace. Another problem, which actually turns out to
be a blessing in disguise, is that, if we fix some specific order
for evaluating the integrals and the finite-dimensional trace $\tr$,
\(naivetrace) gives a finite number
for a much larger class of \psdo s. For example, if we decide to take the
finite-dimensional trace first, then \(naivetrace) will vanish for
any symbol of the type
$s(x,p) = f(x,p) T$, $f: \R^n\times \R^n \to \C$,
$T \in {\frak gl}(M,\C)$ traceless, independently of the the asymptotic
behavior of $s$. Thus, by choosing a particular order of integration,
we can considerably enlarge
the set of symbols yielding a finite answer.
In [Mi-94] it was argued that the right order is to take the
radial momentum integral,
the only potentially divergent one, after the trace and all other
integrals.
The reason why this is the
right thing to do will become abundantly clear from the calculations in
sections~5,~6 and~7.

Quite independent of the above concepts is another
trace that one can define on the space of \psdo s.
This trace, known as the Wodzicki residue [Wo-85, Gu-85, Ad-79, Kr-91, Ma-79],
has many advantages over the one defined by~\(naivetrace).
Thus, consider a \psdo\ $S$ with symbol $s$ having an asymptotic expansion of
the form \(asymptoticexpansion). The  Wodzicki residue of $S$ is defined as
\beq
\res(s)={{1}\over{(2\pi)^n}}\int_{D\times S^{n-1}}
\tr s_{-n}(x,p)\eta(d\eta)^{n-1},
\label{genericwodzicki}
\eeq
where $\eta=p_\mu dx^\mu$ is the canonical one-form, $S^{n-1}$
is the sphere $|p|=\delta$ in momentum space and we are
assuming, as always, that $s_{-n}$ is homogeneous for $|p|\geq \delta$.
Note that \(genericwodzicki) is independent on the radius of the sphere
$\delta$, as long as we assume $s_{-n}$ to be homogeneous outside, and we
could also consider the limit $\limd$ of \(genericwodzicki) as a way
of removing the infrared regulators.
Since we are only considering flat space, expression \(genericwodzicki) reads:
\beq
\res(s)={\delta^n\over{(2\pi)^n}}\int_{|p|=\delta}d\Omega
	\int_{D}d^nx
	\tr s_{-n}(x,p),
\eeq
$d\Omega$ being the angular integration over the sphere $|p|=\delta$.
The residue is a linear functional operator defined on the space of \psdo\
equivalence classes. Notice that it vanishes identically for trace class
operators.

The Wodzicki residue can be used to construct a non-trivial
two-cocycle on the Lie algebra of \psdo s by
\beq
c_R(A,B)=\res([\log|p|,a]_* *b),           \label{genericradul}
\eeq
where $a=\sym A$, $b=\sym B$.
This so-called Radul cocycle [Ra-91a,b]
defines a non-trivial central extension of the
Lie algebra of \psdo s. It also arises in applications of noncommutative
differential geometry~[Co-85, Co-88].

The reader should note that $\log|p|$ is really a singular function
at the origin.
However, the residue is a boundary integral and therefore
independent of the way $\log|p|$ is regularized at the origin.
We also would like to mention that $\log|p|$ does
not have an asymptotic expansion in the sense of eq.~\(asymptoticexpansion).
This does not cause any
problem, however, since only its derivatives appear in the residue.

\section{5. An important lemma: the Radul cocycle as a commutator}

In this section we prove the following identity, to be used in sect.~7:
\beq
     c_R(A,B) = -\intalln ([a,b]_*|_{-n}).
     \label{ourlemma}
\eeq
The importance of the order of the integrals on the r.h.s.~is already
clear at this stage and will become even more obvious after the
calculations. The integrand is not the symbol of a trace class \psdo\ ;
if it were, being a commutator, its trace would vanish. However,
by taking the integrals in the order indicated above, we will be able
to prove that the r.h.s~is well defined (i.e.~independent of the regulators
for the star product) and coincides with the Radul cocycle.
Also, notice that the resemblance of equation \(ourlemma) with a coboundary
$\delta\lambda (A,B) = \lambda([A,B])$ is illusory; the apparent one-cochain
\beq
      \lambda(A) = -\intalln a|_{-n}
\eeq
does not exist since the integral does not converge for a generic element $A$
in the class of \psdo s we are interested in (e.g.~$a=(1+|p|)^{-n}$).

After these words of caution, let us turn to the proof.
Consider two smooth symbols $a$ and $b$, homogeneous of degree $k_a$ and
$k_b$ for $|p|\geq \delta$. We  prove
eq. \(ourlemma) for such symbols --- the complete result follows from
linearity.

Let $N=k_a+k_b+n$. This is the number of $p$-derivatives needed to reach a
symbol of degree $-n$. Using eq. \(stardef) for the star product, the
integrand of the right hand side of eq. \(ourlemma) is written
\beq
	\tr[a,b]_*|_{-n}=\tr{(-i)^N\over N!}
		{\*^Na\over \*p_{\mu_1}\ldots\*p_{\mu_N}}\,
		{\*^Nb\over \*x^{\mu_1}\ldots\*x^{\mu_N}}-(a\leftrightarrow b).
\label{commufirst}
\eeq
No regulating function $\phi$  is needed in \(commufirst) because
we are dealing with a finite sum of smooth functions.
Integration by parts in $x$ is always allowed since the symbols have compact
support. We use this fact to move all $x$-derivatives to $b$ and then identify
the integrand as a total divergence:
\beq
\eqalign{
	&\intalln[a,b]_*|_{-n}\cr
	&\phantom{X}=\intalln{(-i)^N\over N!}\biggl(
	{\*^Na\over \*p_{\mu_1}\ldots\*p_{\mu_N}}\,
		{\*^Nb\over \*x^{\mu_1}\ldots\*x^{\mu_N}}\cr
	&\phantom{XXXXXXXXXXXXXXXX}
	-(-1)^Na{\*^{2N}b\over \*p_{\mu_1}\ldots\*p_{\mu_N}
		\*x^{\mu_1}\ldots\*x^{\mu_N}}\biggr)\cr
	&\phantom{X}=\intalln{i^N\over N!}{\*\over \*p_{\mu_1}}
						\sum_{m=0}^{N-1}(-1)^{m-1}
		{\*^ma\over \*p_{\mu_2}\ldots\*p_{\mu_{m+1}}}\,
		{\*^{2N-m-1}b\over \*p_{\mu_{m+2}}\ldots\*p_{\mu_N}
			\*x^{\mu_1}\ldots\*x^{\mu_N}}\cr
	&\phantom{X}\equiv \intalln {\*\over \*p_\mu}V_\mu.\cr}
\eeq
For later comparison with the Radul cocycle, we have defined quantity
\beq
        V_\mu={i^N\over N!}\tr \sum_{m=0}^{N-1}(-1)^{m-1}
		{\*^ma\over \*p_{\mu_2}\ldots\*p_{\mu_{m+1}}}\,
		{\*^{2N-m-1}b\over \*p_{\mu_{m+2}}\ldots\*p_{\mu_N}
			\*x^{\mu}\*x^{\mu_2}\ldots\*x^{\mu_N}},
                    \label{Vdef}
\eeq
which is homogeneous of degree $(-n+1)$ for $|p|\geq \delta$. Since
the integrand is a total divergence it follows that the integral is
scale invariant, i.e. independent of an ultraviolet cut-off.
Thus, it can be
written as a surface integral that may be pulled back from infinity to
$\delta$:
\beq
	\intalln[a,b]_*|_{-n}={{\delta^{n-2}}\over (2\pi)^n}
		\int_{|p|=\delta}d\Omega \int_D d^nx p_\mu V_\nu
                   \delta^{\mu\nu}.
\eeq
Let us now calculate the Radul cocycle explicitly. The integrand is
\beq
\eqalign{
	&\tr([\log|p|,a]_**b)|_{-n}\cr
	&\phantom{X}=\tr\sum_{q=1}^N{(-i)^N\over q!(N-q)!}\,
		{\*^{N-q}\over \*p_{\mu_{q+1}}\ldots\*p_{\mu_N}}
		\biggl({\*^q\log|p|\over \*p_{\mu_1}\ldots\*p_{\mu_q}}
			{\*^qa\over \*x^{\mu_1}\ldots\*x^{\mu_q}}\biggr)
		{\*^{N-q}b\over \*x^{\mu_{q+1}}\ldots\*x^{\mu_N}}.\cr}
\eeq
Here we need to integrate by parts not only in $x$ but also in $p$, which is
allowed since
\beq
	\int_{D\times S^{n-1}}\biggl({\*\over \*p_\mu}w_\mu\biggr)
	\eta(d\eta)^{n-1}=0\label{dwexact}
\eeq
when $w_\mu$ is homogeneous of degree $(-n+1)$. This follows from the
fact that \(dwexact) is the integral of an exact form
\beq
	\biggl({\*\over \*p_\mu}w_\mu\biggr)\eta(d\eta)^{n-1}=
		d\biggl({1\over n-1}w_\mu dx^\mu\eta(d\eta)^{n-2}\biggr)
	\label{integralform}
\eeq
over a manifold with boundary $\partial (D\times S^{n-1})
\equiv \partial D\times S^{n-1}$ where
$\omega_\mu$ vanishes due to the assumed spatial boundary conditions.

In one dimension the $p$-integral reduces to a sum over
$S^0\!=\!\{\pm \delta\}$.
Eq.\ \(integralform) does not apply in this case, but since the derivative
of a homogeneous function of degree zero vanishes, eq.\ \(dwexact) holds
trivially and formal partial integration is allowed.
We may thus shift all the $p$-derivatives except one from $\log|p|$
and use ${\*\log|p|\over \*p_\mu}={\delta^{\mu\nu}p_\nu\over {p^2}}$ to obtain
\beq
\eqalign{
	&{1\over (2\pi)^n}
        \int_{D\times S^{n-1}}\tr([\log|p|,a]_**b)|_{-n}\eta(d\eta)^{n-1}\cr
	&\phantom{XXX}={1\over (2\pi)^n}\int_{D\times S^{n-1}}
		\eta(d\eta)^{n-1}\tr
		i^N\sum_{q=1}^N\sum_{m=0}^{q-1}{(-1)^{q-1}\over q!(N-q)!}
		{q-1\choose m}\cr
	&\phantom{XXXXXX}\times\,{\delta^{\mu_1\mu}p_\mu\over {p^2}}
		{\*^ma\over \*p_{\mu_2}\ldots\*p_{\mu_{m+1}}}\,
		{\*^{2N-m-1}b\over \*p_{\mu_{m+2}}\ldots\*p_{\mu_N}
			\*x^{\mu_1}\ldots\*x^{\mu_N}}.\cr}
\eeq
Interchanging the order of summation and performing the sum over $q$,
\beq
	\sum_{q=m+1}^N(-1)^{q-1}{N\choose q}{q-1\choose m}=(-1)^m,
\eeq
brings the result
\beq
	\res([\log|p|,a]_**b)=
		-{{\delta^{n-2}}\over (2\pi)^n}
		\int_{|p|=\delta}d\Omega \int_Dd^nx p_\mu
                  V_\nu\delta^{\mu\nu},
\eeq
with $V_\mu$ as in \(Vdef). This proves the lemma. For later purposes, notice
that one can even take the limit $\limd$ in all the equations above,
effectively removing the infrared cut-off from the picture.

Notice that any term in the asymptotic expansion of $[a,b]_*$ after tracing
over ${\frak gl}(M,\C)$ and integrating over $x$ can be written as a total
derivative in $p$. Therefore, any term of degree less than $-n$ vanishes upon
integration over $p$ because of the good ultraviolet asymptotic behavior.
For the term of degree $-n$, on the other hand, the integral becomes
scale invariant instead of having the naive logarithmic
divergence. This we think is at the very heart of the nature of anomalies;
they are neither genuinely ultraviolet nor infrared, but exactly what
is in between.

\section{6. The embedding of $\glpsdo$ in $\glres$ in three dimensions}

Having discussed the basic properties of \psdo s, we are now in a position to
describe the subalgebra of \psdo s in $\glres$, which we denote by
$\glpsdo$. From now on, we shall work in
three dimensions only but it should be clear how to generalize the results to
an arbitrary number of dimensions. As mentioned in sect.~2 we
consider the case of Weyl fermions with an extra ${\frak su}(N)$ degree
of freedom so that ${\cal H} ={\rm L}^2(\R^3) \otimes \C^2_{\rm spin}
\otimes \C^N_{\rm color}$. As before, we shall restrict ourselves to symbols
with compact support $D$ in the variable $x$.  The spin algebra is generated by
the usual Pauli matrices $\sigma^\mu$ ($\mu=1,2,3$ are space indices).

Since the energy of a free Weyl fermion is given by
$E=- i \sigma^\mu\partial_\mu$\footnote{$^\dagger$}{Strictly speaking,
$E$ reverses the
chirality of the spinor. However, for our purposes, we can assume
the existence of a fixed isomorphism between the two chiralities.},
the symbol associated to the sign of the energy is
$\sym(\sE) = {{\sigma^\mu p_\mu}\over{|p|}}$.
As such, this symbol is singular and requires an infrared regularization.
Even if we will never need it
explicitly, one way to regularize a symbol of this kind is to introduce a
function $\phi$ similar to the one used in sect.~4, except
that now $\phi(t)=0$ for $t<\delta/2$, $\phi(t)=1$ for $t>\delta$, and to set:
\beq
       \sym(\sE) = \phi(|p|) {{\sigma^\mu p_\mu}\over{|p|}} \equiv
       \eps. \label{symsign}
\eeq

We can now look for the conditions under which a \psdo\  describes an element
of $\glres$, i.e. has a good second quantization.
Let $A$ be a \psdo\ acting on ${\cal H}$ with symbol
\beq
        a(x,p) \approx \sum_{k\leq m} a_k(x,p), \label{asymptotic}
\eeq
{}From sect.~2, eq. \(defofglres), we must require that $A$ be bounded and
$[\sE, A]$ be HS.
Specializing the considerations of sect.~4 to the $n=3$ case, we must
require for $a$ first of all that $m=0$ and secondly the HS condition
\beq
        \ord([\eps, a]_*)\leq-2. \label{HSsymbol}
\eeq
The most general symbol satisfying these requirements is given by
the asymptotic expansion
\beq
       a(x,p) \approx \sum_{k\leq 0} a_k(x,p),
          \label{asymptoticmeqzero}
\eeq
with
\beq\eqalign{
        a_0(x,p)&=\alpha_0(x,p)+\tilde{\alpha}_0(x,p)\eps, \cr
        a_{-1}(x,p)&={i\over 2}\eps\eps^{\mu}{{\*}\over {\* x^\mu}}
           (\alpha_0(x,p)+\tilde{\alpha}_0(x,p)\eps)
           +\alpha_{-1}(x,p)+\tilde{\alpha}_{-1}(x,p)\eps,\cr
        a_k(x,p)&\hbox{ ~arbitrary} \quad\hbox{for}
          \quad k\leq -2,\cr} \label{explicit}
\eeq
where the expression $\eps^\mu$ denotes
the derivative of the symbol $\eps$ with respect to $p_\mu$ and $\alpha_0$,
$\tilde\alpha_0$, $\alpha_{-1}$ $\tilde\alpha_{-1}$ are four
smooth symbols, homogeneous of
degree $0$ and $-1$ and proportional to the identity matrix in spin space.

To verify that \(explicit) is the most general solution of \(HSsymbol),
expand the commutator $[\eps, a]_*$  and impose that the contributions of
terms of degree $0$ and $-1$ vanishes. This requires:
\beq\eqalign{
               [\eps, a_0] &= 0,\cr
               [\eps, a_{-1}] &= i\eps^\mu{\*\over {\* x^\mu}} a_0,\cr}
               \label{conditionHS}
\eeq
where the commutators in \(conditionHS) are ordinary commutators and we are
considering solutions for $|p|\geq\delta$.
The first of \(conditionHS) has solution
$a_0(x,p)=\alpha_0(x,p)+\tilde{\alpha}_0(x,p)\eps$ since the only $2\times2$
matrices that commute with $\eps$ are the identity and $\eps$ itself.
Plugging this solution into
the second of \(conditionHS), we see that it determines only the component
of $a_{-1}$ that anticommutes with $\eps$.
If we write $a_{-1}=a_{-1}^C + a_{-1}^A$, for the commuting and anticommuting
component respectively, we obtain: $a_{-1}^A =
        {i\over 2}\eps\eps^{\mu}{{\*}\over {\* x^\mu}}
        (\alpha_0(x,p)+\tilde{\alpha}_0(x,p)\eps)$, whereas the commuting part
is given by the most general solution $a_{-1}^C = \alpha_{-1}(x,p)+
\tilde{\alpha}_{-1}(x,p)\eps$.
There are no further requirements on
the star commutator and, therefore, terms of order $\leq -2$ are arbitrary.
This completes the proof of \(explicit). We will in fact never need the
explicit solutions \(explicit) but only use the properties \(conditionHS);
\(explicit) being given for completeness only.
A final remark to be made is that if we tried to solve the second of
\(conditionHS) for $|p|<\delta$ we would have encountered
the problem that, in general, the
equation is not integrable because of the presence of the regulator. However,
these problems do not arise for $|p|\geq \delta$ where
$\eps\equiv {{\sigma^\mu p_\mu}\over{|p|}}$, making the r.h.s.~anticommuting
with $\eps$ and allowing to solve for $a_{-1}^A$.

\section{7. On the cohomology of the Schwinger term in 3+1 dimensions}

In this section, we prove the other main result of our paper:
The Schwinger term for operators in $\glres$, represented by the cocycle
\(schwingerterm)
when restricted to the subalgebra of \psdo s $\glpsdo$
\beq
     \st = \intall ([\eps, a]_* *b),
     \label{schwingerpseudoprel}
\eeq
is cohomologically
equivalent to the ``twisted'' Radul cocycle, defined as
\beq
        c_{TR}(A,B) \equiv c_R(\sE A, B) =
	\res([\log|p|,\eps* a]_* *b) =
        \res(\eps*[\log|p|, a]_* *b).\label{twistedradul}
\eeq
Eq.~\(schwingerpseudoprel) should be interpreted as the limit
\beq\eqalign{
	\st =& \limd\intall\sum_{k=0,-1,-2,-3}([\eps, a]_* *b)|_k\cr &
	+\intall([\eps, a]_* *b)|_{\leq-4},\cr}
		\label{schwingerpseudo}
\eeq
where $\delta$ is the infrared regulator introduced in sect.~4. As we will
see below, there is no need for an ultraviolet cut-off because the
potentially divergent terms will turn out to be zero. Also, we denote by
$s|_{\leq -4}$ a smooth \psdo\ (representative) with asymptotic expansion
$\sum_{k \leq -4} s_k$.

The notion of the twisted Radul cocycle was first introduced in [Mi-94].
To check that $c_{TR}$ really is a two-cocycle is straightforward and will
not be done here (see for instance [Mi-94]).
What is not obvious, however, is that,
despite the fact that expression \(schwingerpseudo) is
{\it not} well defined on the equivalence classes of asymptotic expansions of
\psdo s because of the ambiguity of the integral in
the presence of a regulator, its cohomology is still well
defined in the sense that all dependence on the regularization can be written
as an exact piece $\delta\lambda(A,B)$, the
Lie algebra coboundary of a one-cochain $\lambda$ to be specified below.

The equivalence between these two cocycles was shown to hold for a more
restricted class of operators already in [Mi-94] and, subsequently, in [Fe-94]
for another small class of operators. We show here that the equivalence is
in fact true for all \psdo s  in $\glres$. All previous results follow
straightforwardly from this one. Also, our proof keeps careful track of all
the regulators and allows us to settle some unresolved issues in the previous
literature.

What we will prove is that, for any two operators
$A$ and $B$ in $\glpsdo$ defined through their asymptotic expansions of
the form given in eqs.~\(asymptoticmeqzero) and \(conditionHS),
the following relation
holds:
\beq
        \st = \delta\lambda(A,B) + c_{TR}(A,B). \label{bigdeal}
\eeq
Here $c_{S}$ and $c_{TR}$ are defined as in~\(schwingerpseudo) and
\(twistedradul)
and
\beq
        \lambda(A) = \limd\intdelta(\eps*a)|_{-3} +
        \intall (\eps*a)|_{\leq -4}.\label{coboundaryexplicit}
\eeq
The proof proceeds as follows:
Because of the associativity of the star product, the following relation
between
asymptotic expansions holds true:
\beq
        [\eps, a]_* *b \approx \eps* [a,b]_* + [\eps * b, a]_*.
         \label{associativestar}
\eeq
Now consider the asymptotic expansion of the l.h.s.~in terms of the asymptotic
expansions of $a$ and $b$. The terms of degree $0$ and $-1$ do not appear
because $[\sE, A]$ is a HS operator and $B$ is bounded. The terms of degree
$-2$ and $-3$ can readily be worked out:
\beq\eqalign{
       [\eps, a]_* *b |_{-2} &= \big([\eps,a_{-2}] - i\eps^\mu {\*\over
{\*x^\mu}}
       a_{-1} - {1\over 2}\eps^{\mu\nu}{{\* ^2}\over {\*x^\mu\*x^\nu}}
a_0\big)b_0, \cr
       [\eps, a]_* *b |_{-3} &=\big([\eps,a_{-3}] - i\eps^\mu {\*\over
{\*x^\mu}}
       a_{-2} - {1\over 2}\eps^{\mu\nu}{{\* ^2}\over {\*x^\mu \*x^\nu}} a_{-1}
       +{i\over 6}\eps^{\mu\nu\rho}
		{{\* ^3}\over {\*x^\mu\*x^\nu\*x^\rho}}a_0\big)b_0\cr
        & +\big([\eps,a_{-2}] - i\eps^\mu {\*\over {\*x^\mu}}
       a_{-1} - {1\over 2}\eps^{\mu\nu}
		{{\* ^2}\over {\*x^\mu \*x^\nu}} a_0\big)b_{-1}\cr
       &-i{\*\over {\*p_\rho}} \big([\eps,a_{-2}] - i\eps^\mu {\*\over
{\*x^\mu}}
       a_{-1} - {1\over 2}\eps^{\mu\nu}{{\* ^2}\over {\*x^\mu \*x^\nu}}
a_0\big)
        {\*\over {\*x^\rho}}b_0. \cr} \label{minustwothree}
\eeq
Consider the
function of $p$  arising by taking the finite-dimensional trace and the
integral over the compact domain $D$ of $x$ for both terms in the expansion
\(minustwothree):
\beq\eqalign{
        F_{-2}(p) &= \int_D d^3x \tr [\eps, a]_* *b |_{-2},\cr
        F_{-3}(p) &= \int_D d^3x \tr [\eps, a]_* *b |_{-3}.\cr}
\eeq
The crucial fact is that these two functions {\it vanish}
outside the sphere $|p|=\delta$.
For example, in the case of $F_{-2}$, one can check that neither $a_{-2}$ nor
$a_{-1}^C$ survives the finite-dimensional trace and that the remaining
terms combine to
\beq
        F_{-2}(p)={1\over 2}\int_D d^3x\tr(\eps^\mu\eps\eps^\nu -
\eps^{\mu\nu})
        a_0 {{\* ^2}\over {\*x^\mu \*x^\nu}} b_0,
\eeq
which is zero for $|p|\geq \delta$ because of the identity
\beq
    \eps^{\mu\nu}+\eps\eps^{\mu\nu}\eps+
    \eps^\mu\eps^\nu\eps+\eps^\nu\eps^\mu\eps =0,
\eeq
following by taking two $p$ derivatives of $\eps\equiv\eps\eps\eps$.
In a similar way, the reader can check that also $F_{-3}(p)$ vanishes for
$|p|\geq \delta$.

Using these results, one can restrict the integration over $p$ to the region
$|p|\leq\delta$ for the first four terms ($k=0$, $-1$, $-2$ and
$-3$) in the asymptotic expansion.
(Obviously, the behavior of these functions
inside the sphere depends on the regulator.)
Now note that the integral over $|p|\leq\delta$ of any smooth symbol
of degree $k=0$, $-1$ or $-2$ vanishes as we let $\delta$ go to zero:
\beq
      \limd\intdelta (s_{k}) = 0 \quad\hbox{for}\quad k=0,-1,-2.
      \label{vanishingstuff}
\eeq
Using this fact in  \(schwingerpseudo), one can therefore write:
\beq
\eqalign{
        \st=& \limd\intdelta([\eps, a]_* *b |_{-3})\cr
             	+&\intall ([\eps, a]_* *b |_{\leq -4}).\cr}
\eeq
Using the property \(vanishingstuff) also on the r.h.s.~of
\(associativestar), and comparing with the definition~\(coboundaryexplicit)
we obtain:
\beq\eqalign{
        \st = \delta\lambda(A,B) +& \limd\intdelta
                          ([\eps*b,a]_*|_{-3})\cr +& \intall
                          ([\eps*b,a]_*|_{\leq -4}.\cr}
\eeq
Eq.~\(bigdeal) then follows directly from the results proven in the sect.~5:
\beq
        \limd \intdelta ([\eps*b,a]_*|_{-3}) = -c_{R}(\sE B,A)
        \equiv c_{TR}(A,B)
\eeq
and
\beq
        \intall ([\eps*b,a]_*|_{\leq -4}) = 0.
\eeq

This completes the proof of \(bigdeal). We remark once again that its
importance relies not only on the fact that it relates two seemingly
independent cocycles for the whole space $\glpsdo$ but also
in the fact that it shows that the cohomology of the Schwinger term is well
defined in terms of \psdo s, all the dependence on the regulators being swept
into a coboundary.

\section{8. Conclusions}

In this paper we have shown how to relate two seemingly unrelated concepts
such as the Schwinger term arising in second quantization and the Radul
cocycle.
There are numerous applications, some of which have
already appeared in the literature, that relate directly to our general
theorem. We simply quote some of them.
To begin with, one can indeed reproduce the extension arising in
affine Kac--Moody algebras from the quantization of maps from $S^1$ to
a simple Lie algebra~[Ka-85] and in fact generalize these results from
multiplicative operators to \psdo s. Even more interesting is the
three-dimensional case, which we have discussed at length. If one uses \psdo s
as regularizing
counterterms for higher-dimensional current algebras, as recently proposed
by Mickelsson [Mi-93, Mi-94], one can reproduce the extension arising
in the gauge commutation relations for anomalous chiral gauge theories
directly from the normal-ordered regulated gauge transformations. Other
higher-dimensional current algebras, like the one proposed by us [Ce-94]
also admit such a regularization [Fe-94]. Work is in progress in trying to
understand the possible representation theory for these algebras and we
hope to return on the subject in a later publication.

\section{Acknowledgments.}

We would like to thank J. Mickelsson for having interested us in the
subject and for many crucial discussions. G.F. would also like to
thank E. Langmann for useful conversations.

The work of A.W. was supported in part by funds administered by the Royal
Swedish Academy of Sciences.
\vfill\eject

\section{References}

\item{[Ad-79]}	M. Adler, Invent.~Math.~50 (1979) 219.
\item{[Ba-71]} 	K.~Bardakci and M.B.~Halpern, Phys.~Rev.~D3 (1971) 2493.
\item{[Ba-93]}	I.~Bakas, K.~Khesinad ans E.~Kiritis,
		Commun.~Math.~Phys.~151 (1993) 233.
\item{[Ca-94]} A.L.~Carey and M.K.~Murray, {\tt hep-th/9408141}
\item{[Ce-94]} 	M.~Cederwall, G.~Ferretti, B.E.W.~Nilsson and A.~Westerberg,
               	Nucl.~Phys.~B424 (1994) 97.\hfill\break{\tt hep-th/9401027}
\item{[Co-85]} A.~Connes, Publ.~Math.~IHES 63 (1985) 257.
\item{[Co-88]} A.~Connes, Commun.~Math.~Phys.~117 (1988) 117.
\item{[Fa-84a]} L.~Faddeev, Phys.~Lett.~B145 (1984) 81.
\item{[Fa-84b]} L.~Faddeev and S.~Shatashvili,
		Theor.~Math.~Phys.~60 (1984) 770.
\item{[Fe-94]} 	G.~Ferretti, in ``Proceedings of the G\"ursey Memorial
               	Conference I on Strings and Symmetries'', Istanbul, Turkey
               	(1994), to appear. {\tt hep-th/9406177}
\item{[Fu-90]} K.~Fujii and M.~Tanaka, Commun.~Math.~Phys.~129 (1990) 267.
\item{[Gu-85]} 	V.~Guillemin, Adv.~Math.~55 (1985) 131.
\item{[H\"o-85]}L.~H\"ormander, {\it The Analysis of Partial Differential
		Operators III}, Springer-Verlag, Berlin (1985).
\item{[Ka-67]} 	V.G.~Kac, Funct.~Anal.~Appl.~1 (1967) 328.
\item{[Ka-85]}  V.G.~Kac and D.H.~Peterson, in ``Proceedings of the Summer
                School on Complete Integrable Systems'', Montreal,
		Canada (1985).
\item{[Ka-90]} V.G.~Kac {\it Infinite Dimensional Lie Algebras, 3.~ed.},
               	Cambridge University Press, Cambridge U.K (1990).
\item{[Ki-76]} A.A.~Kirillov, {\it Elements of the Theory of Representations},
               	Springer-Verlag, New York (1976).
\item{[Kr-91]} 	O.S.~Kravchenko and B.A.~Khesin,
		Funct.~Anal.~Appl.~25 (1991) 83.
\item{[La-89]} 	H.~Lawson and M.L.~Michelsohn, {\it Spin Geometry} Princeton
               		University Press, Princeton (1989).
\item{[La-91]} E.~Langmann, in ``Topological and Geometrical Methods in Field
                	Theory'', Turku, Finland (1991).
\item{[La-94a]}E.~Langmann, Commun.~Math.~Phys.~162 (1994) 1.
\item{[La-94b]} E.~Langmann and J.~Mickelsson, Stockholm Royal
                Inst.~Tech.~preprint (1994). \hfill\break{\tt hep-th/9407193}
\item{[Lu-76]} 	L.E.~Lundberg, Commun.~Math.~Phys.~50 (1976) 103.
\item{[Ma-79]} Y.I.~Manin, J.~Sov.~Math.~11 (1979) 1.
\item{[Mi-83]} 	J.~Mickelsson, Lett.~Math.~Phys.~7 (1983) 45.
\item{[Mi-88]} J.~Mickelsson and S.G.~Rajeev,
		Commun.~Math.~Phys.~116 (1988) 365.
\item{[Mi-89]} J.~Mickelsson, {\it Current Algebras and Groups},
               		Plenum Press, New York, (1989).
\item{[Mi-93]} 	J.~Mickelsson, in ``Proceedings of the XXII Int.~Conf.~on
                Differential Geometric Methods in Theoretical Physics'',
               	Ixtapa, Mexico (1993). {\tt hep-th/9311170}
\item{[Mi-94]} 	J.~Mickelsson, Stockholm Royal Inst.~Tech.~preprint (1994).
               		{\tt hep-th/9404093}
\item{[Mo-67]} 	R.V.~Moody, Bull.~Amer.~Math.~Soc.~73 (1967) 217.
\item{[Pi-87]} D.~Pickrell, J.~Funct.~Anal.~70 (1987) 323.
\item{[Pi-89]} D.~Pickrell, Commun.~Math.~Phys.~123 (1989) 617.
\item{[Pr-86]} 	A.~Pressley and G.~Segal, {\it Loop Groups},
		Oxford University Press, Oxford (1986).
\item{[Ra-91a]} A.O.~Radul, Phys.~Lett.~B265 (1991) 86.
\item{[Ra-91b]} A.O.~Radul, Funct.~Anal.~Appl.~25 (1991) 25.
\item{[Sc-59]} 	J.~Schwinger, Phys.~Rev.~Lett.~3 (1959) 296.
\item{[Si-79]} B.~Simon, {\it Trace ideals and their applications},
                Cambridge University Press, Cambridge, (1979).
\item{[Ta-81]} M.E.~Taylor, {\it Pseudodifferential Operators}, Princeton
		University Press, Princeton NJ (1981).
\item{[Va-93]} J.C.~V\'arilly and J.M.~Garcia-Bond\'\i a,
		J.~Geom.~Phys.~4 (1993) 223.
\item{[Wo-85]} 	M.~Wodzicki, in ``K-theory, arithmetic and geometry'',
		Lecture notes in mathematics 1289,
		Springer-Verlag, Berlin (1985).

\bye